\documentclass[prd, twocolumn,nofootinbib,10pt]{revtex4-1}
\date{\today}

\usepackage{epsfig}
\usepackage{amsmath}
\usepackage{amsfonts}
\usepackage{graphicx}
\usepackage{dsfont}
\usepackage[german, english]{babel}
\usepackage{amssymb}
\usepackage{ifthen}
\usepackage{ulem}
\usepackage{color}

\newcommand{\re}[1]{(\ref{#1})}
\newcommand{\ii}{\mathrm{i}}

\graphicspath{{images/}}

\RequirePackage[colorlinks=true,allcolors=blue,
unicode=true,hypertexnames=false]{hyperref}
\hypersetup{pdfstartview={XYZ null null 1.25}}

\begin{document}

\title{Boundary states with elevated critical temperatures in Bardeen-Cooper-Schrieffer superconductors}

\author{Albert Samoilenka}
\thanks{albsam@kth.se}
 \affiliation{Department of Physics, Royal Institute of Technology, SE-106 91 Stockholm, Sweden}
 \author{Egor Babaev}
\affiliation{Department of Physics, Royal Institute of Technology, SE-106 91 Stockholm, Sweden}

\begin{abstract}
Bardeen-Cooper-Schrieffer (BCS) theory describes a superconducting transition as a single critical point where the gap function or, equivalently, the order parameter vanishes uniformly in the entire system. We demonstrate that in superconductors described by standard BCS models, the superconducting gap survives near the sample boundaries at higher temperatures than superconductivity in the bulk. Therefore, conventional superconductors have multiple critical points associated with separate phase transitions at the boundary and in the bulk. We show this by revising the Caroli-De Gennes-Matricon theory of a superconductor-vacuum boundary and finding inhomogeneous solutions of the BCS gap equation near the boundary, which asymptotically decay in the bulk. This is demonstrated for a BCS model of almost free fermions and for lattice fermions in a tight-binding approximation. The analytical results are confirmed by numerical solutions of the microscopic model. The existence of these boundary states can manifest itself as discrepancies between the critical temperatures observed in calorimetry and transport probes.
\end{abstract}

\maketitle

\section{Introduction}
The behavior of the gap function near a boundary of a BCS superconductor \cite{Bardeen1957a} is one of the most basic questions in superconductivity. The problem was widely believed to be solved after the works of Caroli, De Gennes and Matricon (CdGM)  \cite{deGennes_Boundary, CdGM_french, CdGM_Coherence, deGennes_superconductivity}. These works derived the famous boundary conditions for the order parameter \cite{deGennes_Boundary, CdGM_french} using linearized gap equations. The conclusion was reached that the gap at the boundary should vanish at the same temperature as the gap in the bulk, implying that the presence of a boundary in a large system does not alter the conclusions of BCS theory regarding the critical temperature associated with the disappearance of resistivity. For the simplest vacuum-superconductor surface the CdGM boundary condition reads
\begin{equation}\label{boundary_zero}
\vec{n} \cdot \nabla \Delta = 0
\end{equation}
where $\vec{n}$ is the unit vector normal to  the boundary and $\Delta$ is the order parameter. This boundary condition is a cornerstone concept for the interpretation of standard surface probes of a superconductor such as Scanning Tunneling Microscopy.

While the standard textbook picture may leave the impression that this is extremely well theoretically and experimentally understood physics, we
show below that the actual structure of the superconductor-vacuum boundary is principally different. First, let us emphasize that, as CdGM acknowledge themselves, \cite{deGennes_Boundary, CdGM_french} the condition \re{boundary_zero} breaks down in a thin region near the boundary. Moreover, the studies in Refs.~\cite{deGennes_Boundary, CdGM_french} were carried out only under the assumption of a superconducting bulk. However, equality of the critical temperatures for the bulk and the boundary is not a priori guaranteed.
Below, we carry out a more general analysis than CdGM and show that the conclusions in the works \cite{deGennes_Boundary, CdGM_french} are not correct. Instead, even in the simplest BCS model, near a boundary there is a strong increase of the order parameter which in general takes place on a macroscopic  length scale. Hence a superconductor forms a superconducting boundary state. The tail of this state extends inside the bulk over a large length scale 
which makes the CdGM boundary condition incorrect. The consequence of the gap behavior near the boundary that we find is that the superconducting state
gets multiple critical temperatures: above the temperature at which the bulk of the system becomes normal, the boundary retains superconductivity.

This effect is consistent with several experimentally observed facts. First, it is a commonly known fact that resistivity, diamagnetic response and thermodynamic measurements do not in general give the same value for critical temperature $T_c$. By contrast, a commonly observed situation is that a weak superconducting response occurs at slightly higher temperature than a specific-heat jump. Usually this is attributed to the effect of inhomogeneities giving some areas of the sample higher $T_c$. However, the interpretation of these measurements in terms of possible boundary superconductivity in systems with normal bulk has also been voiced \cite{lortz2006origin, janod1993split, butera1988high}. Secondly, possible superconductivity of boundaries in systems with normal bulk was previously invoked in connection with the increase of critical temperatures observed in granular elemental superconductors \cite{deutscher1973granular, cohen1968superconductivity, cohen1967strong}. The basis of this interpretation is that the effect is observed even for relatively large and relatively well-connected granules. The theoretical mechanisms suggested for both cases are (i) the conjecture by V.L.~Ginzburg \cite{ginzburg1964surface, naugle1973evidence} that a surface may modify phonons, and (ii) the conjecture of different chemical composition of the surface. In these scenarios the appearance of boundary states is attributed to a modification of the surface, and hence to the appearance of a different effective Hamiltonian with larger coupling constants near the boundaries, which to the best of our knowledge was not demonstrated at the ab-initio level. Another example of a situation in which superconductivity survives at elevated
temperatures only on the boundary while the bulk is normal was found recently for a pair-density-wave state \cite{Mats_FFLO_SurfaceStates}. However, the mechanism discussed in Ref.~\cite{Mats_FFLO_SurfaceStates} doesn't hold for conventional superconductors. We demonstrate below that more robust superconductivity appears on a clean surface of a BCS superconductor quite generically, and does not require the conjecture of increased coupling constants.

\section{Boundary superconductivity in the model of almost free fermions with attractive interaction}

Let us first recap the standard calculation of the critical temperature for a  superconductor. This can be done for example in Bogoluibov-de Gennes (BdG) formalism \cite{deGennes_Boundary, deGennes_superconductivity, CdGM_french, CdGM_Coherence}, in Gorkov formalism \cite{gorkov1960pis} or directly from a path-integral formulation of the model. Since the transition is of second order, the order parameter should be small near $T_c$. Hence the self-consistency gap equation can be expanded in powers of $\Delta$. To first order this yields \cite{deGennes_Boundary}:
\begin{equation}\label{lin_gap_eq}
\frac{1}{V} \Delta(\textbf{r}) = \int d\textbf{r}' K(\textbf{r}, \textbf{r}') \Delta(\textbf{r}')
\end{equation}

\begin{equation}\label{kernel}
K(\textbf{r}, \textbf{r}') = \int d\textbf{k} d\textbf{k}' F_{\textbf{k}, \textbf{k}'} w_\textbf{k}^*(\textbf{r}) w_{\textbf{k}'}^*(\textbf{r}) w_\textbf{k}(\textbf{r}') w_{\textbf{k}'}(\textbf{r}')
\end{equation}

\begin{equation}\label{F}
F_{\textbf{k}, \textbf{k}'} = T \sum_{\omega} \frac{1}{\xi_\textbf{k} - \ii \omega} \frac{1}{\xi_{\textbf{k}'} + \ii \omega} = \frac{1 - n_F(\xi_\textbf{k}) - n_F(\xi_{\textbf{k}'})}{\xi_\textbf{k} + \xi_{\textbf{k}'}}
\end{equation}
where $\omega = 2 \pi T (\nu + \frac12)$, $\xi_\textbf{k} = E_\textbf{k} -\mu$, $\mu = \frac{k_F^2}{2 m}$, $n_F$ is the Fermi-Dirac distribution and the $w_\textbf{k}$ are single-electron wave functions with eigenvalues $E_\textbf{k}$.

In the bulk of a superconductor, single-electron wave functions can be approximated as plane waves:
\begin{equation}
w_\textbf{k} = \frac{1}{\sqrt{2 \pi}} e^{\ii \textbf{k} \cdot \textbf{r}},\ \ \ E_\textbf{k} = \frac{\textbf{k}^2}{2 m}
\end{equation}

with $\int_{-\infty}^{\infty} w_\textbf{k}(\textbf{r}) w_\textbf{k}^*(\textbf{r}') d\textbf{k} = \delta(\textbf{r} - \textbf{r}')$.
Then in one dimension, in momentum space the linear gap equation \eqref{lin_gap_eq} for $\Delta_k = \int_{-\infty}^{\infty} w_k(x) \Delta(x) dx$
becomes:
\begin{equation}\label{bulk_gap_eq_k}
\frac{1}{V} \Delta_k = D_k \Delta_k
\end{equation}

where $D_k = \frac{1}{2 \pi} \int_{-\infty}^{\infty} F_{k-k', k'} dk'$ is maximal at $k = 0$. For this one-dimensional model we can introduce dimensionless quantities as
\begin{equation}\label{dimensionless_params}
\hat{k} = \frac{k}{2 k_F},\ \ \ \hat{x} = 2 k_F x,\ \ \ \hat{T} = \frac{T}{\mu},\ \ \ \hat{V} = \frac{V}{4 \pi \mu}.
\end{equation}

Hence the only parameters will be $\hat{T}$ and $\hat{V}$. Then $\hat{D}_0 \simeq \ln\frac{c}{\hat{T}}$ with $c = \frac{8 e^\gamma}{\pi}$ for $\hat{T} \to 0$ (namely, $\hat{T} \lesssim 0.1$ or $\hat{V} \lesssim 0.2$). This gives the usual bulk critical temperature:
\begin{equation}\label{Tc1}
\hat{T}_{c1} = c e^{- \frac{1}{\hat{V}}} \ \ \ {\rm for \ the \ bulk \ of \ the \ system.}
\end{equation}

Note that this means that $\Delta_k = \delta(k)$ and hence $\Delta(x) = \text{const}$.

The linear gap equation \re{lin_gap_eq} corresponds to system of linear equations of the form $M_{i j} \Delta_j = 0$, where we for simplicity discretize $\Delta(\textbf{r}),\  K(\textbf{r}, \textbf{r}') \to \Delta_i,\ K_{i,i'}$ and matrix $M = \delta_{i, i'} - K_{i,i'}$. Hence for $T > T_c$ we obtain $\det M \neq 0$, which gives the solution $\Delta = 0$, i.e.\ the normal state. Exactly at $T_c$ $\det M = 0$, which means that there exist a zero eigenvalue of $M$. The corresponding eigenvector gives the configuration of the order parameter up to a multiplicative constant. However, the calculation of $T_c$ is in general nontrivial. While the standard assumption of a homogeneous order-parameter configuration at first glance appears to be reasonable, we show below that it is not accurate. It is important to keep in mind that the conclusion of CdGM about the boundary condition \re{boundary_zero} was derived solely on the basis of the linear gap equation, \cite{deGennes_Boundary, CdGM_french} which does not allow one to assess the configuration of the order parameter at $T < T_c$.

Let us consider the CdGM estimate for the derivative of the order parameter. It is derived in Ref.~\cite{CdGM_french}, equations III.22 -- III.27. To calculate III.26 one needs to estimate $\int d\textbf{r} H(\textbf{r})$, defined by:
\begin{equation}
H(\textbf{r}) \equiv \Delta(\textbf{r}) - \int K^0(\textbf{r},\textbf{r}') \Delta(\textbf{r}') d\textbf{r}'
\end{equation}

where $K^0(\textbf{r},\textbf{r}')$ is defined so that $\int K^0(\textbf{r},\textbf{r}') d\textbf{r}' = 1$ and $K^0(\textbf{r},\textbf{r}') = K^0(\textbf{r}',\textbf{r})$. These properties are used by CdGM to go from equation III.24 to equation III.26. By using the same properties, it is possible to show that the derivation by CdGM of the equation III.26 is not correct. Namely, let us compute the integral:
$
\int d\textbf{r} H(\textbf{r}) = \int d\textbf{r} \Delta(\textbf{r}) - \int d\textbf{r} d\textbf{r}' K^0(\textbf{r},\textbf{r}') \Delta(\textbf{r}') = \int d\textbf{r} \Delta(\textbf{r}) - \int d\textbf{r}' \Delta(\textbf{r}') = 0 \nsim \frac{d\Delta}{dz}
$.
This means that in fact it is not possible to obtain the derivative of the order parameter near the boundary, contrary to the conclusions of Ref.~\cite{CdGM_french}.

Now let us demonstrate that there is a second critical temperature $T_{c2}$ in a semi-infinite superconductor. This effect appears due to the existence of energetically preferred inhomogeneous solutions of the {\it linear} gap equation near the boundary. For simplicity let's consider a one-dimensional semi-infinite superconductor positioned at $x > 0$. The sample is usually modeled to have a periodic potential inside and a finite potential barrier outside. Then single-electron wave functions are given by standing Bloch waves inside and are exponentially suppressed outside. We begin by considering an infinite potential barrier and the almost-free-electron approximation. Thus the single-electron wave function is nullified outside the superconductor. Hence, near the boundary, a single-electron wave function is formed by incoming and reflected electrons, which give standing waves:
\begin{equation}\label{eigen_L}
w_k(x) = 
\begin{cases}
\sqrt{\frac{2}{\pi}} \sin k x, & \text{if}\ x \geq 0\\
0 & \text{if}\ x < 0
\end{cases},\ \ \ 
k > 0,\ \ \ E_k = \frac{k^2}{2 m}.
\end{equation}

We transform to momentum space with $\Delta_k = \int_{0}^{\infty} \phi_k(x) \Delta(x) dx$, where $\phi_k(x) = \sqrt{\frac{2}{\pi}} \cos k x$, which satisfies $\int_{0}^{\infty} \phi_k(x) \phi_k'(x) = \delta(k - k') + \delta(k + k')$. Then the one-dimensional gap equation \eqref{lin_gap_eq} becomes:
\begin{equation}\label{surf_gap_eq_k}
\frac{1}{V} \Delta_k = D_k \Delta_k - \int_{0}^{\infty} A_{k, k'} \Delta_{k'} dk'
\end{equation}

where $A_{k,k'} = \frac{1}{2 \pi} F_{\frac{k + k'}{2},\frac{k - k'}{2}}$ and $D_k = \int_{0}^{\infty} A_{k, k'} dk'$ is the same as in the bulk equation \eqref{bulk_gap_eq_k}. Equation \eqref{surf_gap_eq_k} has to be solved, for a given interaction strength $\hat{V}$, in order to determine whether the boundary has different critical temperature $\hat{T}_{c2}$ than the bulk. Equivalently, one can fix $\hat{T}$ and look for the biggest eigenvalue, which gives $\frac{1}{\hat{V}}$.

\begin{figure}
	\centering
	\includegraphics[width=0.99\linewidth]{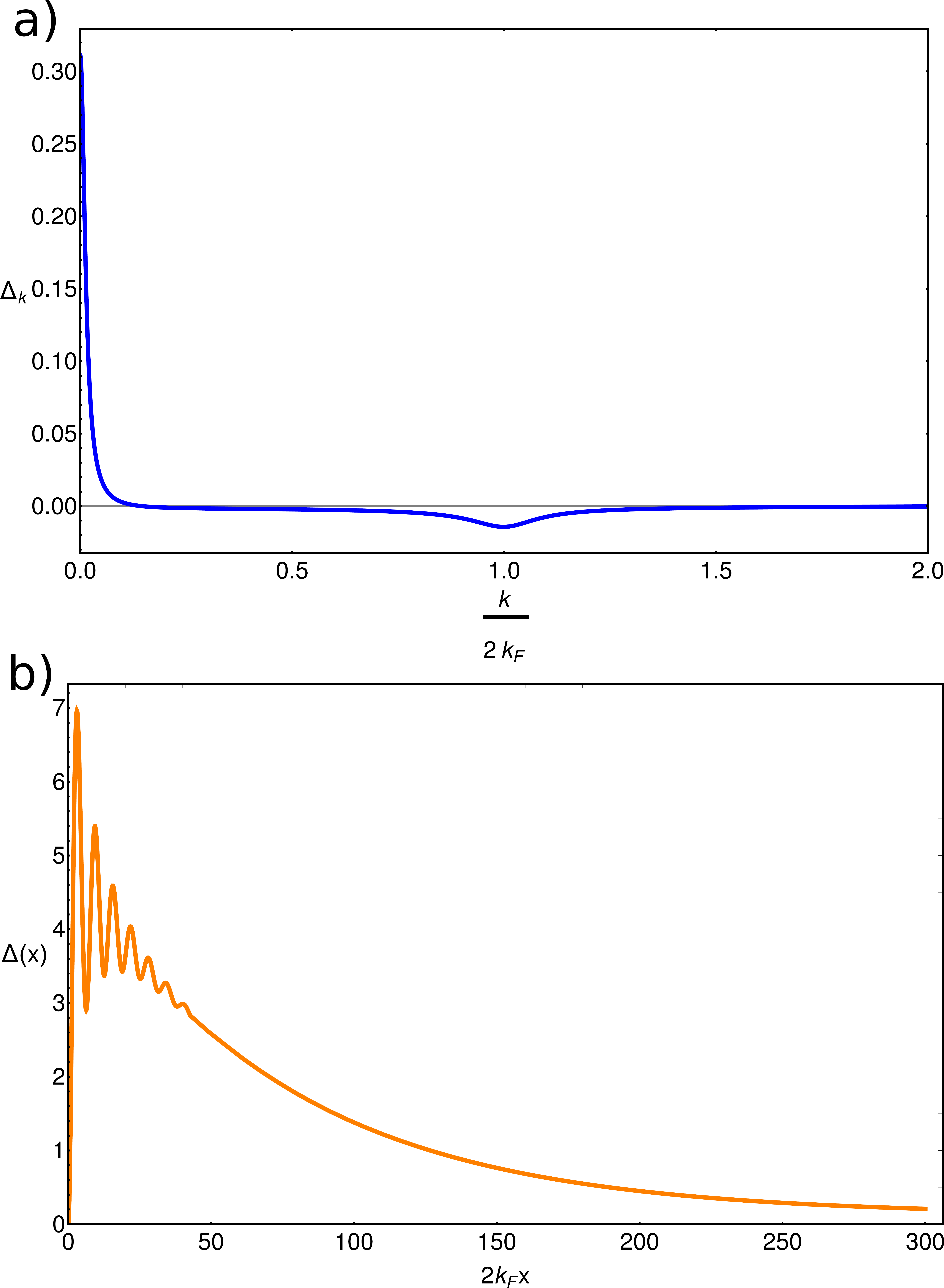}
	\caption{
			The boundary state in the one-dimensional continuous model \eqref{lin_gap_eq} of a half-infinite superconductor at $\hat{T} = 0.05, \hat{V} = 0.219$.
			\textbf{a)} Configuration of the order parameter in momentum space. It shows clear peaks at $k = 0,\ 2k_F$.
			\textbf{b)} The corresponding order parameter in real space with a peak near the boundary at $x = 0$. It decays into bulk at the scale of the bulk coherence length, given by the inverse width of the $k = 0$ peak in panel a.
			Note that near the boundary $\Delta(x)$ has oscillations with frequency $2 k_F$, which decay into the bulk over a dephasing length equal to the inverse width of the $k = 2 k_F$ peak in $\Delta_k$.
	}
	\label{d_plot}
\end{figure}

We obtained solutions numerically by discretizing the integral in Eq.~\eqref{surf_gap_eq_k}. The solutions show that there are superconducting boundary states. A typical example of the configuration of the order parameter in such a state is shown in Fig.~\ref{d_plot}.

Note that the type of configuration shown in Fig.~\ref{d_plot}b is superficially similar to non-superconducting Tamm-Shockley single-electron surface states \cite{Tamm_SS, Shockley_SS}. These have the form of rapidly oscillating standing waves exponentially decaying into the bulk. In contrast to the Tamm-Shockley states, our states are obtained for a superconducting order parameter. Another principal difference is the absence of an assumption of periodic potential in the bulk.

Let us now consider how the boundary state in question is formed. We have three different scales at play here: (i) The shortest one is the \textit{Fermi scale} $\frac{1}{2 k_F}$, which sets the scale of the rapid oscillations of $\Delta(x)$ near the boundary, see Fig.~\ref{d_plot}b. (ii) There are oscillations with slightly different values of $k$. They dephase and disappear over the \textit{dephasing length  scale} $\lambda_d \propto\frac{1}{\delta k_F}$, where $\delta k_F$ is the width of the $k = 2 k_F$ peak in Fig.~\ref{d_plot}a. (iii) There is an increase of the order parameter near the boundary. It serves as a source of superconducting correlations in the bulk. It should asymptotically decay to zero over the {\it bulk  coherence length } $\xi$.

Next, we consider the limit of $T_{c2} \to 0$, or equivalently $V \to 0$. For $x \gg \lambda_d $, the order parameter decays as $\Delta(x) \propto e^{-x / \xi}$. In momentum space this corresponds to
\begin{equation}\label{d_k_0}
\Delta_{k} \propto \delta_\xi(k) \equiv \frac{2}{\pi} \frac{\xi}{1 + \xi^2 k^2}\  \text{for}\  k \ll \delta k_F
\end{equation}

where $\lim\limits_{\xi \to \infty} \int_{0}^{\infty} g(k) \delta_\xi(k) dk = g(0)$. Let us estimate $\xi$. Exponential decay in real space can be obtained by the substitution $k \to \ii/ \xi$. Then, neglecting the second term on the right-hand side of \eqref{surf_gap_eq_k}, we have:
\begin{equation}\label{xi_eq}
\frac{1}{\hat{V}} = \hat{D}_{\ii / \hat{\xi}} \simeq \ln \frac{c}{\hat{T}} + \frac{\alpha}{\hat{T}^2} \frac{1}{\hat{\xi}^2}
\end{equation}
where
\begin{equation}
\frac{\alpha}{\hat{T}} = \int_{0}^{\infty} \hat{A}^2(0,\hat{k}) d\hat{k},\ \text{with}\ \alpha = \frac{7 \zeta(3)}{\pi^2} \simeq 0.85.
\end{equation}
We are looking for the biggest eigenvalue $\frac{1}{V}$ of the integral operator on the right hand side of \eqref{surf_gap_eq_k}. It is possible to find it by the power-iteration method. In other words, we can iteratively apply this operator, until convergence is obtained. In practice, however, starting from Eq.~\eqref{d_k_0} even the first iteration gives a configuration of the  order parameter which is in good agreement with the numerical solution:
\begin{equation}\label{approx_delta}
\Delta_{\hat{k}} \simeq \hat{D}_{0} \delta_{\hat{\xi}}(\hat{k}) - \hat{A}_{\hat{k},0}.
\end{equation}

Now the boundary critical temperature $T_{c2}$ can be estimated by using the   gap equation \eqref{surf_gap_eq_k} for $k = 0$:
\begin{equation}
\frac{1}{\hat{V}} - \ln\frac{c}{\hat{T}} 
= - \int_{0}^{\infty} \hat{A}_{0,\hat{k}} \frac{\Delta_{\hat{k}}}{\Delta_0} d\hat{k} 
\simeq \frac{\pi}{2 \hat{\xi}} \left( \frac{\alpha}{\hat{T} \ln \frac{c}{\hat{T}}} - 1 \right).
\end{equation}

By substituting the coherence length from Eq.~\eqref{xi_eq} we obtain:
\begin{equation}\label{approx_V}
\frac{1}{\hat{V}} - \ln\frac{c}{\hat{T}} 
= \frac{\pi^2}{4 \alpha} \left( \frac{\alpha}{\ln \frac{c}{\hat{T}}} - \hat{T} \right)^2.
\end{equation}

Then using the expression for the bulk critical temperature \eqref{Tc1} we obtain the following expression for the relative boundary-bulk critical temperature difference for $\hat{T}_{ci} \to 0$:
\begin{equation}\label{approx_dT}
\delta \hat{T} \equiv \frac{\hat{T}_{c2} - \hat{T}_{c1}}{\hat{T}_{c1}} \simeq \exp\left( \frac{\pi^2}{4 \alpha} \left( \alpha \hat{V} - c e^{-\frac{1}{\hat{V}}} \right)^2 \right) - 1.
\end{equation}

Note that the effect studied here is very easy to miss in an approximate calculation. For example, $\delta \hat{T} = 0$, if one follows the common procedure and averages the oscillations over the Fermi scale $\frac{1}{k_F}$. This is because these oscillations are encoded in $\hat{A}_{\hat{k},0}$ in \eqref{approx_delta}, which is essential for correct calculation of the difference in the critical temperatures.

To ascertain that we are not overestimating the critical temperature of the boundary states in \eqref{approx_dT} we can use the following inequality for the approximated value of $\frac{1}{\hat{V}}$:
\begin{equation}\label{inequalty_V}
\frac{1}{\hat{V}} = \frac{\int_{0}^{\infty} \hat{D}_{\hat{k}} |\Delta_{\hat{k}}|^2 d\hat{k} -  \int_{0}^{\infty} \Delta_{\hat{k}} \hat{A}_{\hat{k}, \hat{k}'} \Delta_{\hat{k}'} d\hat{k} d\hat{k}'}{\int_{0}^{\infty} |\Delta_{\hat{k}}|^2 d\hat{k}} \leq \frac{1}{\hat{V}_{\text{exact}}}.
\end{equation}

\begin{figure}
	\centering
	\includegraphics[width=0.99\linewidth]{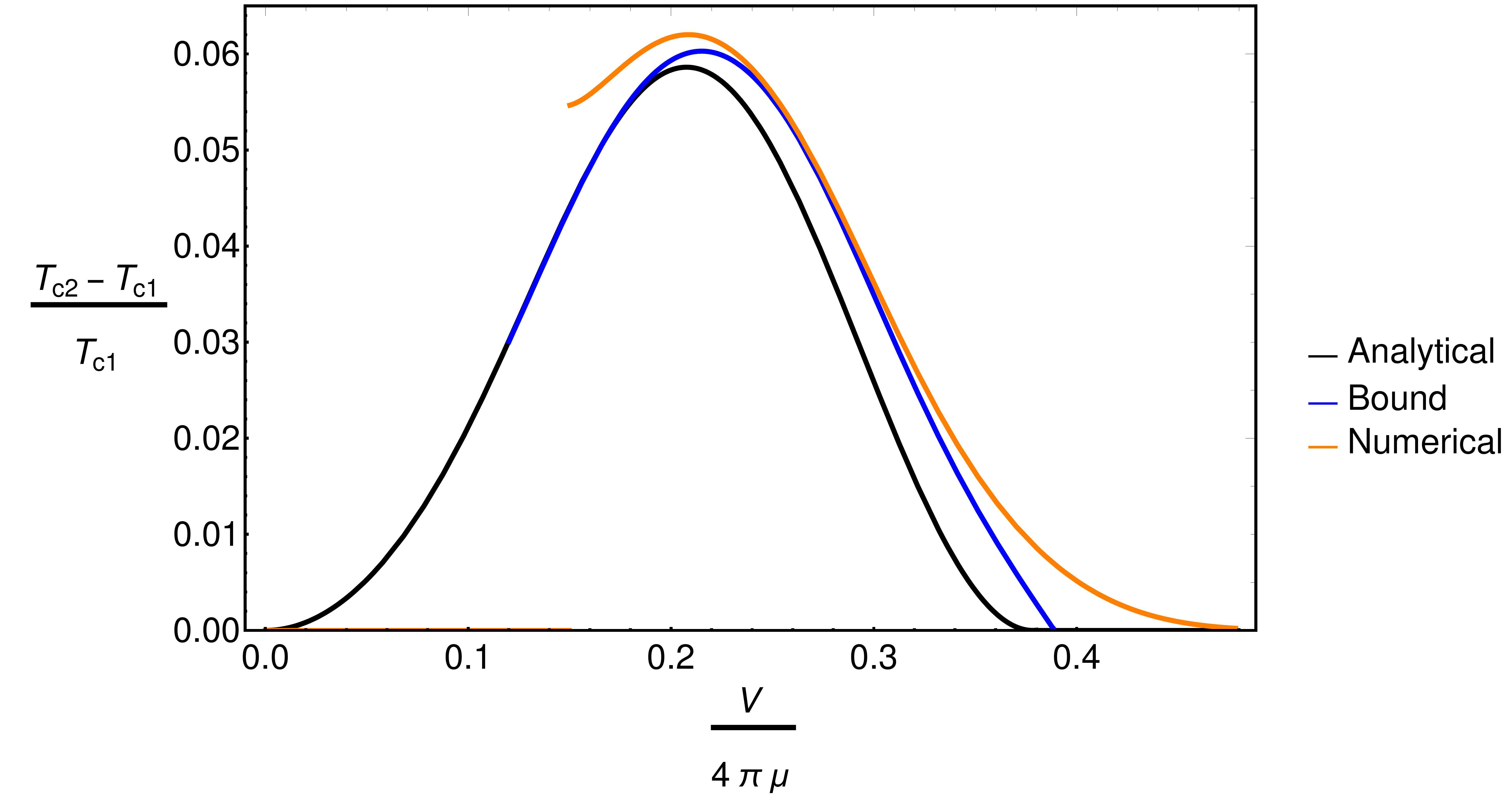}
	\caption{
			The relative difference $\delta T$ of the critical temperatures of the boundary ($T_{c2}$) and of the bulk ($T_{c1}$) as a function of the rescaled interaction strength $\hat{V}$ in the one-dimensional model of almost free electrons with attractive interaction \eqref{surf_gap_eq_k}.
			(Black) analytical approximation \eqref{approx_dT}, which works well for $\hat{V} \lesssim 0.2$ and rather well for $\hat{V} \lesssim 0.4$.
			(Blue) Bound on minimal value of $\delta T$ given by Eq.~\eqref{inequalty_V}, which is based on Eq.~\eqref{approx_dT}.
			Due to the numerical integration, this can be calculated for $\hat{V} \gtrsim 0.1$.
			(Orange) Numerical solution of the discretized equation \eqref{surf_gap_eq_k}. The solution loses accuracy very fast for $\hat{V} \lesssim 0.2$.
			Note that the difference is biggest for intermediate interaction strength and goes to zero in the limit of zero coupling constant and in the strong coupling limit.
			The latter is natural to expect.
			Indeed, deep in the BEC limit one should recover the standard bosonic behavior: suppressed rather than enhanced superconductivity near the boundaries.
	}
	\label{dT_cont}
\end{figure}

Then from Eqs.~\eqref{inequalty_V} and \eqref{bulk_gap_eq_k} we can obtain a lower bound on the critical temperature difference $\delta T$. Combining this result with numerical solutions we obtain the dependence of $\delta T$ on interaction strength $\hat{V}$ shown in Fig.~\ref{dT_cont}.

The physical interpretation of the above derivation is the following: the Cooper pairs near a boundary are made of electrons in a superposition of incoming and reflected waves (i.e.\ a standing wave). This type of pairing wins over the conventional Cooper pairing of electrons with opposite momenta. This means that in order to observe an increase of the critical temperature one should study surfaces or other kinds of interfaces that efficiently reflect  electrons, resulting in the formation of standing waves. An example is a superconductor-to-insulator boundary.

\section{Boundary superconductivity in the tight-binding model of fermions}
In the previous section we demonstrated the existence of boundary states in the simplest model of almost free electrons with weak attractive interaction. In order to make a more general assessment of the presence or absence of this effect, in this section we consider the case of lattice fermions in a tight-binding model. In this section we utilize the Bogoliubov-de-Gennes approach.
In the tight-binding limit the standard Hamiltonian for an s-wave superconductor reads \cite{BdG_Book_Zhu}:
\begin{equation}\label{bdg_ham}
H = \sum_{i, j = 1, \sigma = \pm}^{i, j = N} c^\dagger_{i \sigma} h_{i j} c_{j \sigma} + \sum_{i = 1}^{N}\left( \Delta_{i} c_{i \uparrow}^\dagger c_{i \downarrow}^\dagger + \Delta_{i}^* c_{i \downarrow} c_{i \uparrow}\right)
\end{equation}

where $h_{i j} = - t_{i j} - \mu \delta_{i j}$ and the hopping coefficient is nonzero only for nearest neighbors $t_{\langle i j \rangle} = t$.

First we will consider one-dimensional systems. We have to work with a sufficiently large system size (with number of sites up to $N = 100$) to avoid finite-size effects. Coefficients were set to $t = 1$, $\mu = \frac12$ and $V = 2$.

We iteratively diagonalize the Hamiltonian with Bogoliubov transformations:
\begin{equation}
\begin{aligned}
&c_{i \sigma} = \sum_{n = 1}^{2 N} u_{i\sigma}^n\gamma_n - \sigma v_{i \sigma}^{n*}\gamma_n^\dagger,\\
&c_{i \sigma}^\dagger = \sum_{n = 1}^{2 N} u_{i\sigma}^{n*}\gamma_n^\dagger - \sigma v_{i \sigma}^{n}\gamma_n
\end{aligned}
\end{equation}

And recompute $\Delta_i$ from the self-consistency equation: 
\begin{equation}\label{bdg_gap_eq}
\Delta_{i} = \frac{V}{2}\sum_{n = 1}^{2 N} { u^{n}_{i\uparrow}v^{n*}_{i\downarrow}\tanh\left(\frac{E_n}{2k_B T}\right)}
\end{equation}

\begin{figure}
	\centering
	\includegraphics[width=0.99\linewidth]{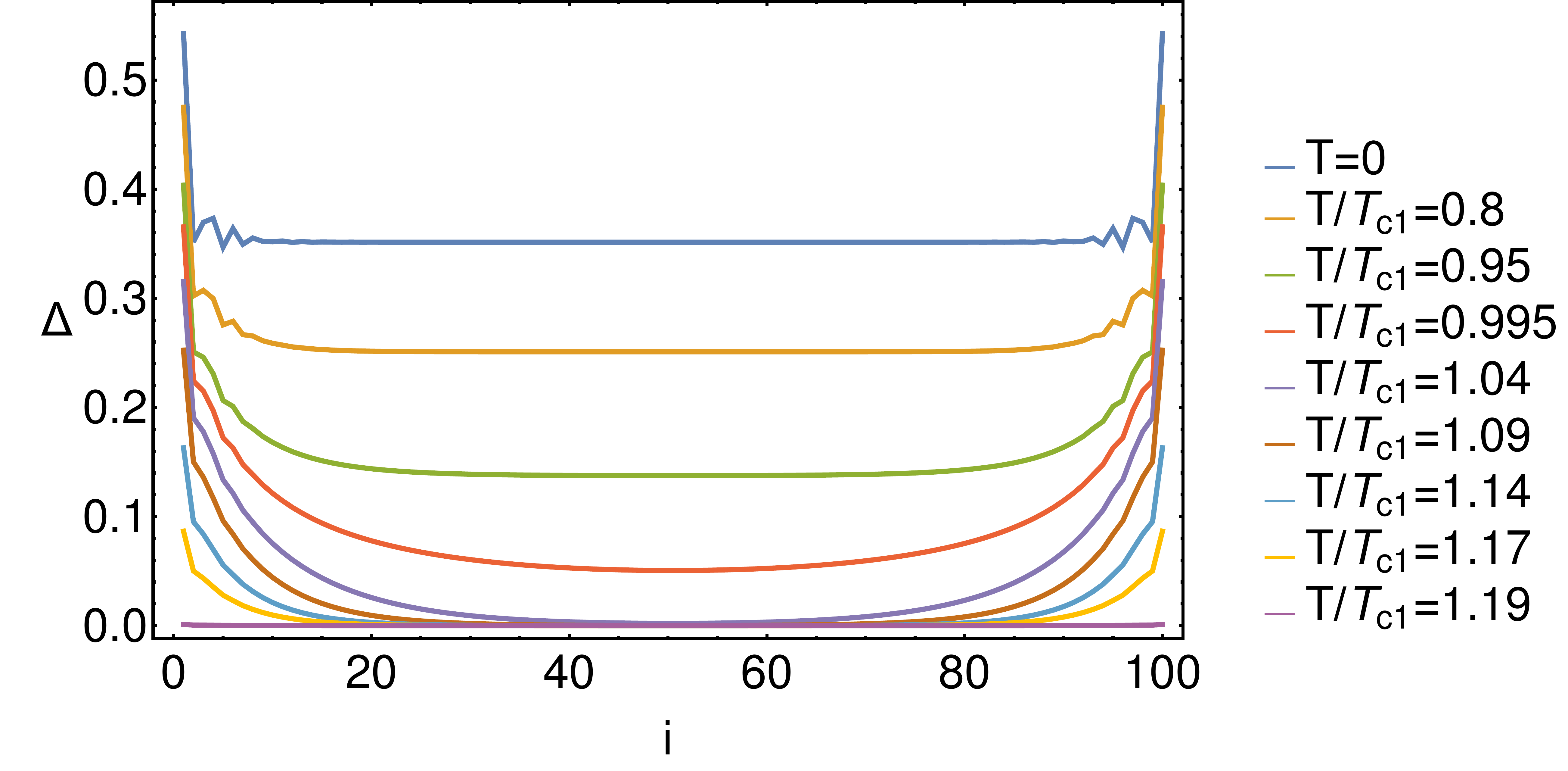}
	\caption{
	Boundary states in one-dimensional tight-binding Bogoliubov-de Gennes model with free boundaries.
	The single-electron wave function vanishes outside the superconductor on the length scale of the tail of a Wannier function, hence outside $\Delta = 0$.
	The spacial configuration of the order parameter $\Delta_i$ is presented for several temperatures below and above 
	the bulk critical temperature $T_{c1}$.
	Observe the gap increase close to the boundary. 
	The boundary state relaxes to a uniform configuration over a length scale corresponding to the bulk coherence length $\xi$.
	The order parameter clearly has nonzero derivative on the boundary in contradiction to the CdGM condition \re{boundary_zero}.
	}
	\label{bdg_d}
\end{figure}

After $\sim 10^3$ iterations we obtain converged states depicted in Fig.~\ref{bdg_d}. These states are global minima and have energy lower than the normal state. Observe that for $T < T_{c1}$ the superconducting gap is uniform in the bulk but has an increase near the boundary. We find that this increase gradually vanishes in the bulk on the length scale of the bulk coherence length $\xi$.

The numerical calculations confirm that the largest extent of the boundary state takes places near the critical temperature associated with the bulk transition $T \simeq T_{c1}$. Indeed, at $T_{c1}$ the bulk coherence length $\xi$ diverges and the boundary state has a power-law decay into the bulk. For $T > T_{c1}$ a superconductor is normal in the bulk. However, in the tight-biding regime we similarly find that the gap remains non-zero near the boundaries above the temperature of the bulk transition. Note that in Fig.~\ref{bdg_d} the order-parameter field varies relatively smoothly since it's defined only on sites, which corresponds to averaging or evaluating at the peaks of the oscillations of Fig.~\ref{d_plot}b. Note that the situation is similar to that for almost free electrons, considered in the previous section. Namely, the free boundary conditions here correspond to perfect reflection of the particles. This allows the formation of boundary states due to the increase of the density of states near the boundaries.

To understand the oscillation of the gap near the boundary, let us now make an analytic approximation for $T < T_{c1}$. It is possible to diagonalize the hopping term in the tight-binding Hamiltonian \re{bdg_ham} using standing waves by performing the following transformation \cite{tight_binding_diagonal_french,tight_binding_diagonal}:
\begin{equation}\label{c_trans}
c_{i \sigma} = \sqrt{\frac{2}{N + 1}} \sum_{n = 1}^{N} \sin\left(k_n i \right) d_{n \sigma}
\end{equation}

where $k_n = \frac{\pi}{N + 1} n$.
In the limit $T \to 0$, the terms that depend on $\Delta$ are diagonalized and we obtain the Hamiltonian:
\begin{equation}
H \simeq - \sum_{n \sigma} \xi_n d_{n \sigma}^\dagger  d_{n \sigma} + \sum_{n} \left( \tilde{\Delta}'_n d_{n +}^\dagger  d_{n -}^\dagger + (\tilde{\Delta}'_n)^* d_{n -}  d_{n +} \right)
\end{equation}

where $\xi_n = 2 t \cos(k_n) + \mu$ and
\begin{equation}
\frac{2}{N + 1} \sum_{i = 1}^{N} \sin(k_{n_1} i) \sin(k_{n_2} i) \Delta_i \simeq \delta_{n_1, n_2} \tilde{\Delta}'_{n_1}.
\end{equation}

Next the solution for $\Delta_i$ can be composed of standing waves and a constant as follows:
\begin{equation}\label{d_i}
\Delta_i = \frac{1}{\sqrt{N + 1}} \sum_{n = 0}^{N} \cos(2 k_n i) \tilde{\Delta}_n.
\end{equation} 

Since a superconductor has a uniform gap in its bulk, for $T \to 0$ the gap is almost constant except a small region near the boundary and hence $\tilde{\Delta}_0 \gg \tilde{\Delta}_{n \neq 0}$ and $\tilde{\Delta}'_n \simeq \langle \Delta \rangle \simeq \frac{\tilde{\Delta}_0}{\sqrt{N + 1}}$, where $\langle \Delta \rangle$ is the average order parameter in the bulk.
The gap equation \re{bdg_gap_eq} gives for $n \neq 0$:
\begin{equation}\label{d_n}
\tilde{\Delta}_n = - \frac{V}{2} \frac{1}{\sqrt{N + 1}} \frac{\langle \Delta \rangle}{\sqrt{\xi_n^2 + \langle \Delta \rangle^2}} \tanh \frac{\sqrt{\xi_n^2 + \langle \Delta \rangle^2}}{2 T}
\end{equation}

where $\langle \Delta \rangle$ is defined through the usual gap equation:
\begin{equation}
1 = \frac{V}{2} \frac{1}{N + 1} \sum_{n = 1}^{N} \frac{\tanh \frac{\sqrt{\xi_n^2 + \langle \Delta \rangle^2}}{2 T}}{\sqrt{\xi_n^2 + \langle \Delta \rangle^2}}
\end{equation}

\begin{figure}
	\centering
	\includegraphics[width=0.99\linewidth]{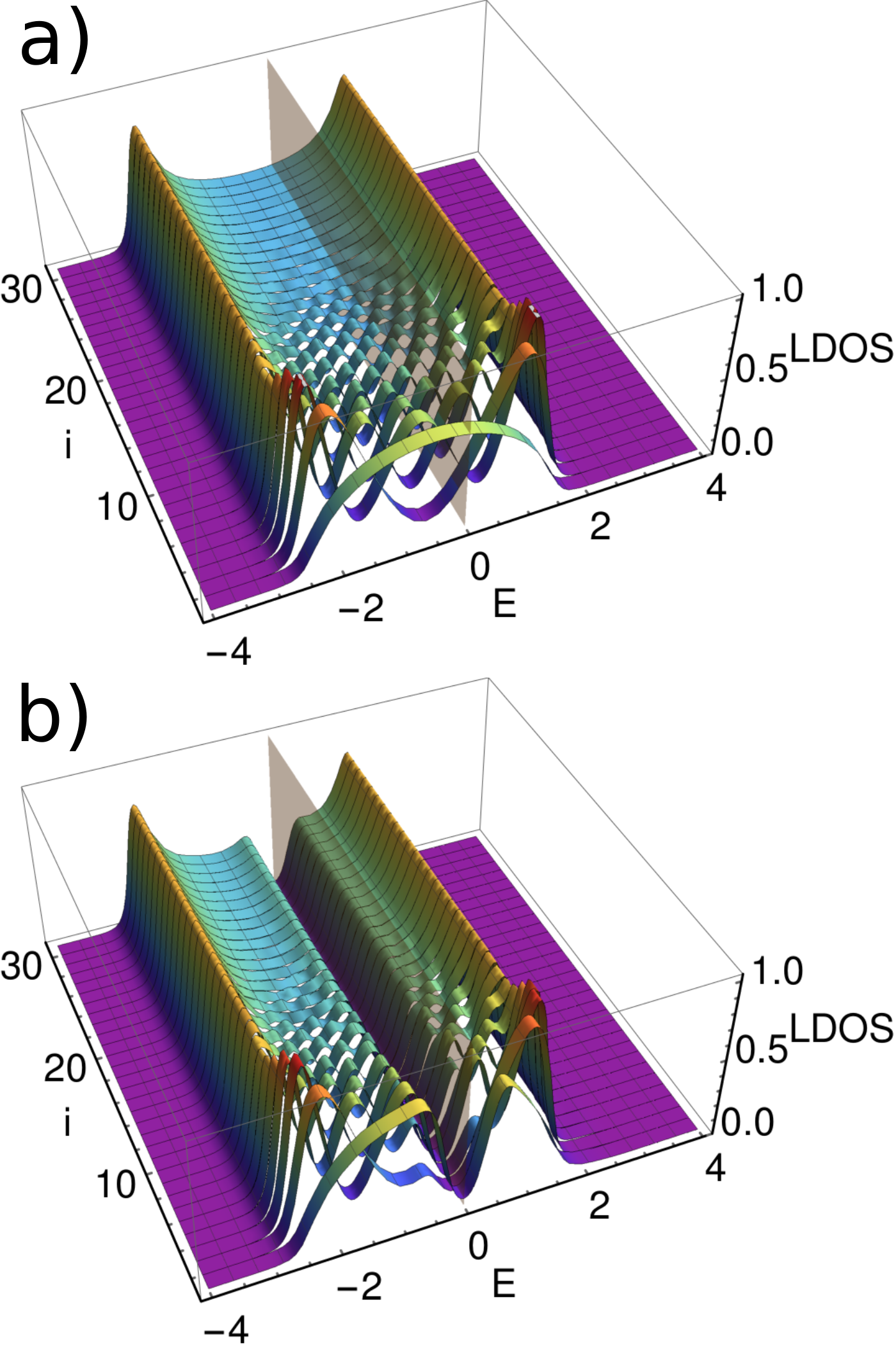}
	\caption{
Local density of states (LDOS) \re{LDOS_eq} as a function of energy $E$ and site number $i$ for the first $30$ sites near a boundary of a chain with $N = 100$ sites at $T = 0.1$. The orange plane corresponds to the Fermi level.
\textbf{a)} Model without interaction between electrons $V = 0$. Note that at a given site $i$ the LDOS is modulated by the square of the eigenstates $\frac{2}{N + 1} (\sin k_n i)^2$ according to \re{LDOS_eq}. Hence the LDOS on the boundary on site $i = 1$ can be up to two times bigger than in the bulk, i.e.\ form an upshoot.
\textbf{b)} The superconducting gap for $V = 2$, showing enhancement on the boundary.
	}
	\label{LDOS}
\end{figure}

Then we use Eq.~\re{d_n} to transform back to real space. By employing Eq.~\re{d_i} we obtain a real-space $\Delta_i$ configuration which is quite close to the numerically obtained solution shown in Fig.~\ref{bdg_d} for $T = 0$. The crucial point in this analysis is that when performing the summation \re{d_i}, essentially, one represents an almost constant $\Delta_i \simeq \langle \Delta \rangle$ order parameter by some set of standing waves. Thus, clearly, oscillations inevitably should appear on the boundaries. In a general context this type of effect is known as the Wilbraham-Gibbs phenomenon \cite{wilbraham1848certain, gibbs1899fourier}. Similarly, the electron density exhibits oscillations and increases near a boundary. Another way to put it is that the Friedel oscillations of electron density are strongest close to the boundary. Most importantly, there are oscillations and a sharp increase in the local density of states (LDOS). These oscillations in the LDOS are presented in Fig.~\ref{LDOS}, calculated using the equation \cite{BdG_Book_Zhu}:
\begin{equation}\label{LDOS_eq}
\rho_i(E) = -\sum_{n} \left[ |u^n_{i \uparrow}|^2 n_F'(E_n - E) + |v^n_{i \downarrow}|^2 n_F'(E_n + E) \right].
\end{equation}
The spikes in the LDOS produce the increase in the gap, which then relaxes into the bulk at the length scale of the bulk coherence length $\xi$.

Although the gap function given by \eqref{d_n} exhibits characteristic oscillations at $2 k_F$ as in the continuous case (Fig.~\ref{d_plot}a), it doesn't give a good approximation for the critical temperature. Therefore, in order to calculate the critical temperature we use the linear gap equation for the tight-binding case. This is just a discrete version of (\ref{lin_gap_eq}, \ref{kernel}, \ref{F}) with the single-electron wave function and energies given by:
\begin{equation}\label{non_periodic}
\text{non periodic:}\ w_n(i) = \sqrt{\frac{2}{N + 1}} \sin(k_n i),\ E_n = 2 t \cos k_n
\end{equation}

\begin{equation}\label{periodic}
\text{periodic:}\ w_n^{(p)}(i) = \frac{1}{\sqrt{N}} e^{-\ii p_n i},\ E_n^{(p)} = 2 t \cos p_n.
\end{equation}

Here $N$ is the number of sites in the corresponding direction, $p_n = \frac{2 \pi n}{N}$ and $k_n$ was defined in \eqref{c_trans}.

Firstly, let's consider the periodic case. We transform to momentum space by $\tilde{\Delta}_n = \sum_{i = 1}^{N} (w_n^{(p)}(i))^* \Delta_x$ with $\sum_{i = 1}^{N} (w_n^{(p)}(i))^* w_{n'}^{(p)}(i) = \delta_{n,n'}$ for $n,n'\in[1,N]$.
Then the one-dimensional linear gap equation becomes:
\begin{equation}\label{per_lin_gap_eq}
\frac{1}{V} \tilde{\Delta}_n = D_n^{(p)} \tilde{\Delta}_n
\end{equation}

where
\begin{equation}\label{per_D_n}
D_n^{(p)} = \frac{1}{N} \sum_{n' = 1}^{N} F^{(p)}_{n - n', n'}.
 \end{equation}

Here $(p)$ means that the Eq.~\eqref{periodic} is used. Equations \eqref{per_lin_gap_eq} and \eqref{per_D_n} are trivially generalized to higher dimensions. The interaction strength $V$ is numerically calculated for given bulk critical temperature $T_{c1}$ and chemical potential $\mu$ from $\frac{1}{V} = D_N^{(p)} = D_0^{(p)}$.

Note that $t$ can be scaled away. Hence, we set  $t = 1$.

Now let us consider the case where in one of the directions the system is not periodic, i.e.\ it has open boundary conditions. Using the transformation \eqref{d_i} to go to momentum space, we obtain the following one-dimensional linear gap equation:
\begin{equation}\label{non_per_lin_gap_eq}
\frac{1}{V} \tilde{\Delta}_n = D_n (\tilde{\Delta}_n + \tilde{\Delta}_{N + 1 - n}) - \sum_{n' = 0}^{N} A_{n, n'} \tilde{\Delta}_{n'}
\end{equation}

where we used Eq.~\eqref{non_periodic}.
$D_n$ and $A_{n, n'}$ are defined as:
\begin{equation}\label{disr_A}
A_{n, n'} = \frac{1}{2 (N + 1)} \left( F_{n + n', n - n'} + F_{n + n' + N + 1, n - n' + N + 1} \right)
\end{equation}

\begin{equation}
D_n =
\begin{cases}
\frac{1}{2} \sum_{n' = 0}^{N} A_{n, n'} \ \text{if}\ n \neq 0 \\
\sum_{n' = 0}^{N} A_{n, n'} \ \text{if}\ n = 0.
\end{cases}
\end{equation}

This equation is easy to generalize to higher dimensions by adding periodic directions. For example, for a two-dimensional system that is non-periodic in $x$ and periodic in $y$, this amounts to replacing $\xi \to E_{n_x} + E^{(p)}_{n_y} - \mu$ and performing the sum $\frac{1}{N_y} \sum_{n_y = 1}^{N_y}$ in \eqref{disr_A}.

\begin{figure}
	\centering
	\includegraphics[width=0.99\linewidth]{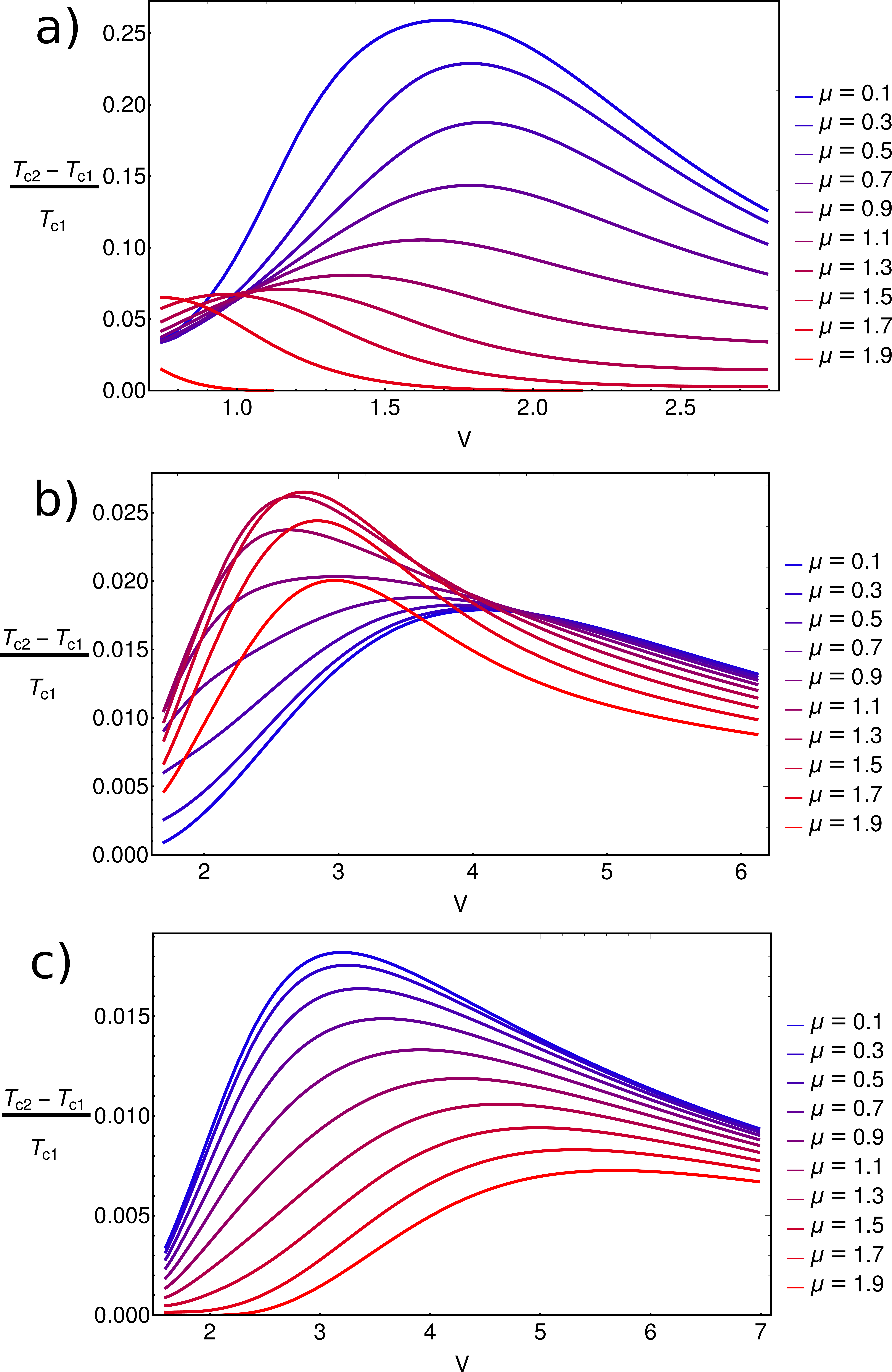}
	\caption{
			The relative difference $\delta T$ between the boundary ($T_{c2}$) and bulk ($T_{c1}$) critical temperatures as a function of interaction strength $V$ for several values of the chemical potential $\mu$. 
			The figure shows the effect in the \textbf{a)}one-, \textbf{b)}two- and \textbf{c)}three-dimensional tight binding models \eqref{bdg_ham}.
			Note that in all cases boundary states are formed and hence $T_{c2} > T_{c1}$.
			The critical temperature difference is maximal for intermediate coupling $V$ and goes to zero for strong and weak interaction limits.
	}
	\label{dT_all}
\end{figure}

One can find $T_{c2}$ by solving the equations with fixed interaction. We followed an equivalent procedure that gives the same solution in a different way. We fixed the boundary critical temperature $T_{c2}$ and chemical potential $\mu$ to various values and numerically solved the right hand side of \eqref{non_per_lin_gap_eq} for the biggest eigenvalue, which equals $\frac{1}{V}$.
Combining this with data from Eq.~\eqref{per_lin_gap_eq}, we obtain the relative difference in critical temperatures for a range of $V$ and $\mu$, see Fig.~\ref{dT_all}.
For this we numerically simulated systems of $N_x = 1000$ and $10000$ in one dimension, $N_x = 1000,\ N_y = 1000$ in two dimensions and $N_x = 1000,\ N_y = 100,\ N_z = 100$ in three dimensions. The results are shown in Fig.~\ref{dT_all}.
The boundary states we find are most prominent in one dimensional systems and the relative critical temperature difference is an order of magnitude smaller in two- and three- dimensional systems.
The reason why the effect is smaller in higher dimensions is the following: when introducing another periodic direction, $A_{n, n'}$ becomes effectively averaged over $[-2 t -\mu, 2 t - \mu]$ in \eqref{non_per_lin_gap_eq}.
This, as can be seen from Fig.~\ref{dT_all}a, will decrease the critical temperature difference.
This leads to the conclusion that superconductivity can survive to substantially higher temperatures in low-dimensional systems.
This includes the edges and corners of three-dimensional cuboidal samples and boundaries with large curvature in general.
That is, these results show that if one increases the temperature in a cuboid sample, then at mean-field level, the following sequence of the transitions will take place: first, the gap will close in bulk, but will remain on the surfaces, increasing temperature further will make the gap vanish on the surfaces, but the gap will remain on the edges of the sample.
Increasing temperature further will result in the gap closure at the edges, but the gap will survive to a higher temperature in the corners of the sample.

\section{Transparency of interfaces}

\begin{figure}
	\centering
	\includegraphics[width=0.99\linewidth]{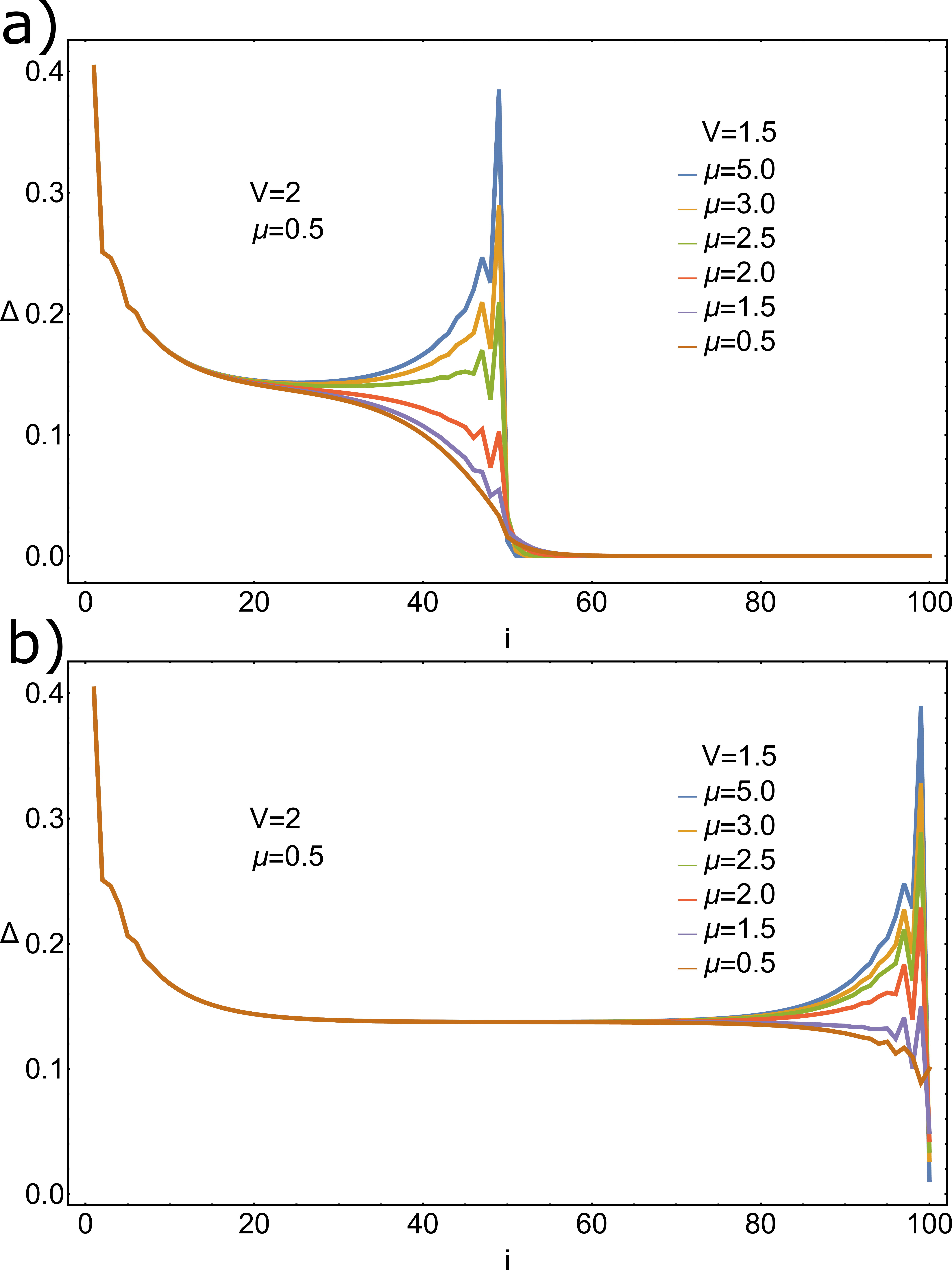}
	\caption{
		Different types of interfaces.
		The spacial configuration of the order parameter $\Delta_i$ in one-dimensional tight-binding Bogoliubov-de Gennes model with free boundaries for $T / T_{c1} = 0.95$ (here $T_{c1}$ is the bulk critical temperature for the parameters used for the left part of the system).
		\textbf{a)}Interface of superconductor ($i < 50$) and non-superconducting material ($i \geq 50$).
		The normal material is modeled here by choosing $V = 1.5$ and different values of chemical potential $\mu$.
		In case of superconductor-normal interface the standard proximity effect behavior is recovered, where the gap is suppressed at superconducting side and a proximity tail with different correlation length is present in the normal side.
		This changes to the gap enhancement behavior when the onsite potential is large and the interface is more reflective.
		\textbf{b)} Superconductor with a different composition of a boundary layer $i = 100$ for which the coupling constant is decreased to $V = 1.5$ and various $\mu$.
		The smaller coupling constant at the interface, even at the scale of one lattice point shown here, can lead to significant suppression rather than enhancement of the gap value close to the surface.
	}
	\label{interface}
\end{figure}

In this section we briefly comment on the implication of our findings for the other types of interfaces.
The Fig.~\ref{interface}a shows an example of a numerical solution for the simplest interface between a superconductor and a normal material obtained numerically in the BdG model.
There we change the onsite potential $\mu$ for the normal half of the system.
When $\mu$ is small the interface is transparent and the standard proximity effect behavior \cite{deGennes_Boundary} is observed.
When increasing the onsite potential, which makes the interface more reflective, the behavior changes gradually and eventually the gap function develops a spike near the interface.
We note that the width of the solution, both in this case and for the boundary state is macroscopic: i.e. it is of the scale of bulk coherence length.
Therefore we believe that the effect may in principle be described in the Ginzburg-Landau or quasiclassical \cite{zaitsev1984quasiclassical,kuprianov1988influence} approaches.
However, a naive quasiclassical approach will not capture the effect: width of the solution is given by coherence length but, the source of enhanced superconducting correlations is at microscopic length scale.
Thus to describe this effect in these approaches a careful treatment of microscopic degrees of freedom is required.
This problem of compatibility of the results with quasiclassical theory will be addressed elsewhere.

The problem of superconductor-normal interface is related to the the problem of different type of superconducting surfaces.
Namely, in a real material a surface can be oxidized, have different chemical composition or different phonons, resulting in weaker coupling constant.
On Fig.~\ref{interface}b we show how depending on the characteristic of the altered boundary layer, the gap can be suppressed or enhanced at a macroscopic length scale away from the boundary.

\section{Conclusions}

In conclusion, we showed that in the BCS model there are several critical temperatures at the mean-field level.
The superconducting gap survives at the faces of a sample up to a higher temperature than in the bulk.
Furthermore, at mean field level it survives to a higher temperature at the edges and even higher at vertices.
This suggests an explanation of the frequently observed disparity between critical temperatures determined by specific-heat and diamagnetic-response probes. A resolution of the phenomena, that we find, in specific-heat measurements would require a homogeneous cuboid sample with a perfect surface.
A number of experiments have explicitly claimed observation of boundary superconductivity above the temperatures where the samples lost superconductivity in the bulk, including hints of such behavior in the specific heat alone \cite{lortz2006origin,janod1993split,butera1988high}. The most advanced experimental analysis of this effect was performed on YBCO. There the conclusion about superconducting surfaces was reached by cutting samples and  observing the same splitting of the phase transitions. In these experiments the difference between $T_{c1}$ and $T_{c2}$ is consistent with our calculations. 
We calculated, using several approaches, $T_{c2}$ for a clean, absolutely reflecting boundary. For a rough surface the gaps will be higher on convex parts, while concave parts will represent weak links, so that a rough surface will not necessarily be able
to have high critical current but still would contribute to the diamagnetic response. Apart from that, the found boundary states are important for interface superconductors and interpretation of Scanning Tunneling Microscopy probes. The solution for the boundary gap and the sequence of phase transitions can be directly studied in fermionic ultracold atoms in a box potential \cite{box1, shin2007tomographic, schirotzek2008determination}. Beyond mean-field approximation, the bulk transition will be in the 3D XY universality class and the surface transition will be of Berezinskii-Kosterlitz-Thouless type, which can also be experimentally probed.
On the other hand we show that going beyond the homogeneous BCS model and adding even a small layer with weaker coupling, i.e. due to oxidation, different chemical composition, or different electron-phonon coupling can lead to a suppression rather than enhancement of the gap.

{\it Note added:} 
During the review process we were pointed to the article \cite{giamarchi1990onset} that analyzed a related model and arrived at the opposite conclusion that there is no boundary superconductivity in a system with an attractive BCS fermion-fermion interaction. Another opposite  conclusion in \cite{giamarchi1990onset} is the behavior of the order parameter near a boundary when the bulk of the system is superconducting. According to \cite{giamarchi1990onset}, for a BCS system with attractive interaction the gap exhibits only Friedel oscillations. In other words, it is enhanced only at a microscopic length scale, while averaging over microscopic length scales the order parameter stays constant at the length scale of the coherence length. Next, only by going to a different, specifically modified model was it possible to achieve a relative $T_c$ difference of the order $10^{-3}$. We investigated the discrepancy by performing momentum space calculations, similar to \cite{giamarchi1990onset}, and obtained a numerical solution for the unmodified model in one dimension.
Based on that calculation, our conclusions are opposite: we do observe a higher critical temperature near the boundaries and the enhancement of the average gap at a macroscopic length scale. Our interpretation of the reason for the discrepancy is the following: The conclusion in \cite{giamarchi1990onset} was obtained numerically. We note that discretizing the integral in the linear gap equation is a subtle problem -- the lower $\hat{V}$ is the harder it is to obtain precise results. In our one-dimensional case we had to go up to $6 \times 10^4$ points to obtain somewhat exact results for $\hat{V} \simeq 0.2$.

\textbf{Acknowledgements}

We thank Mats Barkman, Emil Blomquist, Andrea Benfenati, Filipp N.\ Rybakov, Attila Geresdi, Wolfgang Ketterle and Vladimir Krasnov for discussions. We especially thank Vadim Grinenko for pointing out the experiments claiming the observation of two critical temperatures after this work was published as an eprint. The work was supported by the Swedish Research Council Grants No.\ 642-2013-7837 and No.\ VR2016-06122, 2018-0365 and the Goran Gustafsson Foundation for Research in  Natural Sciences and Medicine.

\bibliographystyle{apsrev4-1}
%\bibliography{references}

%

\end{document}